\documentclass[12pt]{article}
\usepackage{amssymb}
\usepackage[dvips]{graphicx}

\begin{document}

\parskip=0.3cm
\begin{titlepage}

\hfill \vbox{\hbox{DFPD 02/TH/1}
\hbox{January 2002}}

\vskip 0.5cm

\centerline{\bf Pomeron exchange and $t$-dependence of the scattering amplitude}

\vskip 0.3cm

\centerline{F. Paccanoni}

\vskip 0.1cm

\centerline{\sl Dipartimento di Fisica, Universit\`a di Padova,}
\centerline{\sl Istituto Nazionale di Fisica Nucleare, Sezione di 
Padova,}
\centerline{\sl via F. Marzolo 8, I-35131 Padova, Italy}

\vskip 0.1cm

\begin{abstract}
Constraints on the $t$-dependence of the hadronic scattering amplitude 
at asymptotic energies are derived by considering the exchange of the 
Pomeron, as a Regge pole, between off-shell gluons. Covariant 
reggeization ensures pure spin $\alpha$ exchange, where $\alpha$ is the 
Regge trajectory of the Pomeron. The structure of the amplitude, as a 
function of $t$, has been derived without a specific choice for the 
partonic wave functions of the hadrons. New terms appear, with respect 
to the standard approach, and allow to describe non trivial properties 
of the diffraction cone in agreement with experimental data, as shown in 
a specific example.

\end{abstract}

\vskip 0.1cm

\vfill

\hrule

$ \begin{array}{ll}
\mbox{{\it email address:}} &
\mbox{PACCANONI~@PADOVA.INFN.IT}
\end{array}
$

\end{titlepage}
\eject
\newpage

\section{Introduction}
Field theoretical descriptions of hadronic two-body processes at 
large $s$ and small $|t|$, based on the idea that strong interactions 
are mediated by colored gluons exchanged between color singlet states, 
have a long history. The identification of the Pomeron with two, or 
more, gluon exchanges~\cite{LN}, gave rise to the BFKL 
equation~\cite{BL,BFKL} whose asymptotic solution is the so-called 
"Lipatov Pomeron". The Born approximation to the Pomeron, used also 
recently in many calculations of high energy processes, developed in a 
conformal field theory, at least in the leading-log approximation.
\vskip 0.3cm
At large momentum transfer squared, the perturbative approach will give 
sensible results for the observables of any hadronic exclusive process.
Non perturbative effects can however be important at small $|t|$, surely 
they are present in the Born approximation, and 
corrections to the Pomeron pole in perturbative QCD are larger than 
expected in the next-to-leading approximation~\cite{FL}. It could be 
interesting, at this stage, to consider a phenomenological model 
where, first, the Pomeron is exchanged as a Regge pole between off-shell 
gluons and, then, off-shell gluons couple to a color singlet, e.g. a 
$q\bar{q}$ state. A more sophisticated model, where quark and many-gluon 
components of the Pomeron are considered, would be more 
realistic, but far more involved. The quark component, however, can be 
taken into account, in the simplest case, and this will be done in the 
following.
\vskip 0.3cm
Covariant reggeization~\cite{MDS,OT,GJ} provide us with a differential 
technique that generates reggeized scattering amplitudes free from kinematical
singularities and satisfying automatically factorization. The most 
important properties of this approach were its direct relation with 
Toller poles~\cite{TO} and the possibility to include easily spins of 
the particles coupled to the Pomeron ensuring, for the latter, pure spin 
$\alpha$ exchange where $\alpha$ is its Regge trajectory~\cite{OT,DA}.
It could seem, at first sight, that this method does not have any 
predictive power since the couplings depend on a large number (five for 
the gluonic component)
of unknown functions of $t$. As will be seen later, constraints on the 
coupling of two off-shell gluons to "white" hadrons, yielding even C 
exchange, simplify noticeably this picture~\cite{BL,GS,LR}.
\vskip 0.3cm
Unless a specific choice of the partonic wave functions for the 
interacting hadrons and of the Pomeron couplings to off-shell gluons
is made, only formal properties of the scattering 
amplitude can be derived with the aforementioned technique. It turns 
out, however, that these properties are interesting enough to justify 
the general treatment of Section {\bf 2}, where conditions on the scattering 
amplitude, frequently required on phenomenological grounds, are derived 
with a limited number of assumptions. In Section {\bf 3}, a model will be 
considered for the Pomeron as a double Regge pole and formal constraints 
on the $t$ dependence of the amplitude will be derived. In this step 
a possible treatment of the non-perturbative region is proposed  
based on a suitable regularization procedure. In Section {\bf 4}, the new amplitude 
is compared with the standard Regge formalism by 
considering a particular example, proton-antiproton scattering at high 
energy. Generalizations to $\gamma$ induced processes, like photoproduction, 
will be also touched upon. Concluding remarks appear in Section {\bf 5}.

\section{Covariant reggeization}
 
Consider first Fig.~1, the bare Pomeron in our approach. 
$\gamma, \delta, \nu$ and $\rho$ are gluons and the zigzag line the 
Pomeron Regge pole. Let $Q=k' +r/2, P=k\,-r/2, \Delta=r$ and 
$\Delta^2=t<0$. If $(P\cdot Q)$ is large, the contribution for this graph 
can be written in the form~\cite{OT,GJ}
\begin{equation}
{\cal M}_g= 
\frac{2^{\alpha}\Gamma(\alpha+3/2)}{\sqrt{\pi}\Gamma(\alpha+1)}\xi_+M_g
\label{z1}
\end{equation}
where
\begin{equation}
M_g= C^+_{\nu\rho}C^+_{\gamma\delta} 
\frac{\partial}{\partial P_{\alpha_1}}\frac{\partial}{\partial P_{\alpha_2}}
\frac{\partial}{\partial Q_{\beta_1}}\frac{\partial}{\partial 
Q_{\beta_2}} (P\cdot Q)^{\alpha}
\label{z2}
\end{equation}
and $\xi_+=\exp(-i\pi\alpha/2)/\sin(\pi\alpha/2)$ is the signature;  
$\alpha$ is the Pomeron trajectory $(\alpha=\alpha(t))$. 
Expressions (\ref{z1}, \ref{z2}) ensure pure 
spin $\alpha$ exchange but, since "external" particles are off-shell 
gluons, the usual mass shell 
conditions cannot be used after all the derivatives have been done.
Hence, if the reduced Regge coupling $C^+_{\nu\rho}$ is written in the 
general form
\begin{displaymath}
C^+_{\nu\rho}=g_1 P_{\alpha_1}P_{\alpha_2}P_{\nu}P_{\rho}+g_2 
P_{\alpha_1}P_{\alpha_2}g_{\nu\rho}+
\end{displaymath}
\begin{equation}
+g_3 g_{\alpha_2\nu}P_{\alpha_1}P_{\rho}+g_4 g_{\alpha_1\rho}P_{\alpha_2}P_{\nu}+
g_5 g_{\alpha_1\rho}g_{\alpha_2\nu}
\label{z3}
\end{equation}
the coefficients $g_i$ depend a priori on all the invariants $t, k^2$ 
and$(k-r)^2$: $g_i\equiv g_i(t,k^2,(k-r)^2)$. An expression, analogous to 
(\ref{z3}) holds for the upper vertex $C^+_{\gamma\delta}$.

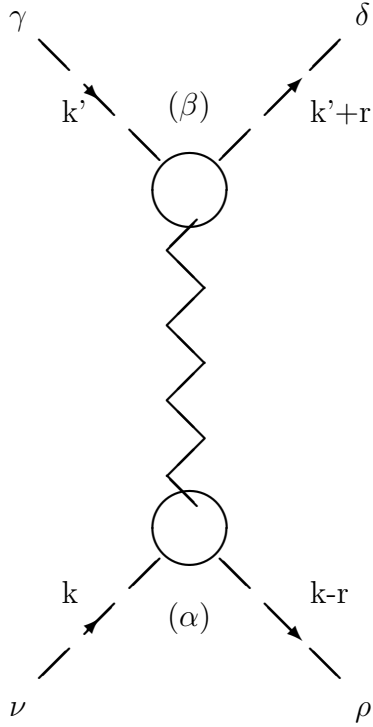
\begin{figure}[bt]
\setlength{\unitlength}{1mm}
\begin{center}
\begin{picture}(60,110)
\thicklines
\multiput(10,100)(6,-6){3}{\line(1,-1){4}}
\multiput(34,84)(6,6){3}{\line(1,1){4}}
\multiput(10,15)(6,6){3}{\line(1,1){4}}
\multiput(34,31)(6,-6){3}{\line(1,-1){4}}
\put(30,80){\circle{10}}
\put(30,35){\circle{10}}
\put(27,72){\line(1,1){4}}
\put(27,42){\line(1,-1){4}}
\multiput(27,72)(0,-10){3}{\line(1,-1){5}}
\multiput(27,62)(0,-10){3}{\line(1,1){5}}
\put(15,95){\vector(1,-1){3}}
\put(42,92){\vector(1,1){3}}
\put(15,20){\vector(1,1){3}}
\put(42,23){\vector(1,-1){3}}
\put(6,102){$\gamma$}
\put(52,102){$\delta$}
\put(6,10){$\nu$}
\put(52,10){$\rho$}
\put(13,89){k'}
\put(13,25){k}
\put(46,89){k'+r}
\put(46,25){k-r}
\put(27,90){$(\beta)$}
\put(27,22){$(\alpha)$}
\end{picture}
\end{center}
\caption[]{\small Scattering of off-shell gluons $g^*+g^*\to g^*+g^*$ with 
Pomeron exchange.}
\label{fig1}
\end{figure}

A first simplification, if the Pomeron is coupled to hadrons,
$a$ and $b$ with masses $m_a$ and $m_b$,  as in Fig.~2, comes 
from Ward identities. The gluon-particle amplitude, call it 
$f_{\nu\rho}$, vanishes at the lowest order when saturated with $k^{\nu}$ 
or $(k-r)^{\rho}$~\cite{BL,LI} and, neglecting non leading terms of the 
form $r^{\nu} f_{\nu\rho}$, only the terms with coefficients $g_2$ and 
$g_5$ in (\ref{z3}) give a non zero contribution. Notice that all the 
terms, having origin from the derivatives in eq.~(\ref{z2}), are leading; 
at the end they will give rise to contributions going as $(P\cdot Q)^{\alpha}$.
The term proportional to $g_2$, that is usually neglected, represents a 
new and interesting feature of the scattering amplitude.

\begin{figure}[tb]
\setlength{\unitlength}{1mm}
\begin{picture}(70,60)
\thicklines
\put(5,10){\line(1,0){15}}
\put(55,10){\line(1,0){15}}
\put(5,50){\line(1,0){15}}
\put(55,50){\line(1,0){15}}
\put(37.5,50){\oval(35,10)}
\put(37.5,10){\oval(35,10)}
\put(25,25){\framebox(25,10){Fig. 1}}
\multiput(30,15)(0,6){2}{\line(0,1){4}}
\multiput(30,35)(0,6){2}{\line(0,1){4}}
\multiput(45,15)(0,6){2}{\line(0,1){4}}
\multiput(45,35)(0,6){2}{\line(0,1){4}}
\put(10,53){$p_a$}
\put(10,13){$p_b$}
\put(64,53){$p_a+r$}
\put(64,13){$p_b-r$}
\put(10,10){\vector(1,0){3}}
\put(60,10){\vector(1,0){3}}
\put(10,50){\vector(1,0){3}}
\put(60,50){\vector(1,0){3}}
\put(30,-5){Fig.~2}
\end{picture}
\hfill
\begin{picture}(75,55)
\thicklines
\put(5,10){\vector(1,0){5}}
\put(65,10){\vector(1,0){5}}
\put(10,10){\line(1,0){5}}
\put(70,10){\line(1,0){5}}
\put(18,10){\oval(6,14)}
\put(62,10){\oval(6,14)}
\put(21,15){\vector(1,0){8}}
\put(28,15){\line(1,0){30}}
\put(21,5){\line(1,0){8}}
\put(59,5){\vector(-1,0){30}}
\put(40,40){\circle{10}}
\put(35,15){\circle*{1}}
\put(45,5){\circle*{1}}
\multiput(35,15)(0,6){4}{\line(0,1){4}}
\multiput(45,5)(0,6){6}{\line(0,1){4}}
\put(35,23){\vector(0,1){3}}
\put(45,25){\vector(0,-1){3}}
\put(8,12){$p_b$}
\put(72,12){$p_b-r$}
\put(32,35){$\nu$}
\put(48,35){$\rho$}
\put(28,27){k}
\put(48,27){k-r}
\put(28,18){$l_1$}
\put(28,1){$l_2$}
\put(35,12){$\sigma$}
\put(45,1){$\tau$}
\put(17,8){V}
\put(61,8){V}
\put(30,-5){Fig.~3}
\end{picture}
\vskip 0.3cm
\caption{\small Hadron scattering with the insertion of the $g^*-g^*$ 
amplitude.}
\label{fig2}
\caption{\small Notation for the formal integration at the lower vertex.}
\label{fig3}
\end{figure}

Consider first, for the sake of simplicity, the scattering hadrons as
$q\bar{q}$ bound states; extrapolation to real mesons or baryons will 
not change the main result of the model.
The formal evaluation of the vertices, for example the lower one shown in 
Fig.~3, can then be done as follows. Momentum-space techniques~\cite{CW} can 
be used by choosing, in the $s\to\infty$ limit, a reference frame where 
the large components of the momenta of the incoming and outgoing 
particles are along the $z$-axis. Hence, setting $\omega=\sqrt{s}/2$, 
from Fig.~2 one has~\footnote{For the four vector $a$, the infinite 
momentum variables are defined as 
$a_{\pm}=a^0\pm a^3$ and $\vec{a}_{\perp}=(a^1, a^2).$}: $p_a=(p_{a+}, 
p_{a-},\vec{p}_{a\perp})= (2\omega, 
m_a^2/(2\omega),0),\;p_b=(m_b^2/(2\omega),2\omega,0)$ and $r= 
(r_+,r_-,\vec{r}_{\perp})$ with
\begin{displaymath}
r_{\pm}=\frac{\pm\,t\,(p_{a\pm}+p_{b\pm})}{p_{a+}p_{b-}-p_{a-}p_{b+}}\simeq 
\pm {\cal O}(t/\omega)
\end{displaymath} 
In the following, the masses and $r_+, r_-$ will be neglected with respect
to $\omega$. In Fig.~3, that represents one of the three 
possible diagrams to be evaluated, the particle with momentum $l_1$ is 
a quark and $l_2$ an antiquark with $l_1-l_2=p$. The leading 
contribution of the lower part of this graph will have a tensor 
structure of the form $l_1^{\sigma}\cdot l_2^{\tau}$, or $l_1^{\nu}\cdot 
l_2^{\rho}$ when the gluon propagators in Feynman gauge are taken into 
account. With the position $l_1=p/2+z,\;\;l_2=-p/2+z$ the integration over 
$d^4k\;d^4z=(1/4)dk_+\;dk_-\;d\vec{k}_{\perp}\;dz_+\;dz_-\;d\vec{z}_{\perp}$
can be formally performed.
\vskip 0.3cm
 Whatever the form of the vertex $V$ and the functions $g_2$ and 
$g_5$ could be, the integrals over $k_+$ and $z_+$, at the lower vertex, 
can be done by closing over the respective poles. The same procedure 
applies to the integrals over $k_-'$ and $z_-'$ at the upper vertex.
Keeping always the leading terms in $\omega$, the integrations over the 
transverse momenta and $z_-$ can be performed implicitly; they involve 
in fact the unknown functions $g_2,\;g_5$ and $V$.  
Only the integrals over the "large" components of $k$ and $k'$, $k_-$ 
and $k_+'$, remain undone. While $|k_-|$ and $|k_+'|$ are large, they 
satisfy the constraint $k_-, k_+' \ll \omega$ and, for simplicity sake, 
are neglected with respect to $\omega$. A more rigorous approach 
would not change sensibly the final form of the amplitude whose 
structure is determined from the model chosen for the Pomeron 
propagator.
\vskip 0.3cm
Let now $X$ be a second rank tensor constructed from the four-momenta 
$k, k'$ and $r$. When the aforesaid integrations have been performed,  
the following correspondences can be established
\begin{equation}
g_2 g_{\nu\rho} X^{\nu\rho} \to V_2^{(b)}(t),\;\;\;\;\;\;g_5 
g_{\alpha_1\nu} g_{\alpha_2\rho}X^{\nu\rho} \to V_5^{(b)}(t)X^{--}
\label{z4}
\end{equation}
\begin{equation}
g_2 g_{\gamma\delta} X^{\gamma\delta} \to V_2^{(a)}(t),\;\;\;\;\;\;g_5 
g_{\beta_1\gamma}g_{\beta_2\delta}X^{\gamma\delta} \to V_5^{(a)}(t) 
X^{++}
\label{z5}
\end{equation}
where $V_{2,5}^{(a,b)}(t)$ are unknown functions of $t$, unless a 
specific choice for the wave functions and $g_2,\;g_5$ is made.
\vskip 0.3cm
For the scattering of identical particles, $a\equiv b$, one gets, from 
the $M_g$ part of the propagator,
\begin{displaymath}
\alpha^2(\alpha-1)^2\left[\frac{1}{\omega^2}V_2^2(t)(P\cdot Q)^{\alpha}+
\frac{1}{4}V_2(t)V_5(t)(k_-^2+k'_+{}^2)(P\cdot 
Q)^{\alpha-2} \right.
\end{displaymath}
\begin{equation}
\left. +\frac{\omega^2}{4}V_5^2(t)(P\cdot Q)^{\alpha-2} \right]
\label{z6}
\end{equation}
where
\begin{equation}
P\cdot Q \simeq \frac{k_- k_+'}{2}.
\label{z7}
\end{equation} 
is the large variable. As far as the 
practical evaluation of the integrals in (\ref{z4},\ref{z5}) is 
concerned, while the term with $g_5$ is standard, the calculation of the 
term with $g_2$ will be far from trivial and gives rise, in perturbative 
QCD, to singularities both in the infrared and in the ultraviolet 
regions. The latter singularities can be avoided if the function $g_2$ 
helps to make the integrals convergent. 
The integrals over $k_-$ and $k_+'$ factorize and are both of the form
\begin{displaymath}
\left(\frac{1}{2}\right)^{\alpha-m} \int^{\omega/\rho}\,dk_-\,k_-^{\alpha 
+n-m}
\end{displaymath}    
where $n, m$ are integers and the scale $\rho, \rho>1,$ has been chosen the 
same at both vertices for simplicity sake.
The integration of eq.~(\ref{z6}), with the position $\rho^2=s_0/4$, 
gives
\begin{displaymath}
\frac{\alpha^2(\alpha-1)^2}{2^{\alpha-1}}\left( 
\frac{\omega^2}{\rho^2}\right)^{\alpha} \left[ 
\frac{V_2(t)}{\rho(\alpha+1)}+\frac{\rho V_5(t)}{\alpha-1}\right]^2=
\end{displaymath}
\begin{equation}
\frac{\alpha^2}{2^{\alpha-1}(\alpha+1)^2}\left(\frac{s}{s_0}\right)^{\alpha}
\left[\frac{2(\alpha-1)}{\sqrt{s_0}}V_2(t)+\frac{\sqrt{s_0}(\alpha+ 
1)}{2} V_5(t) \right]^2.
\label{z8}
\end{equation}
Hence the model requires the presence of a large scale $s_0$.
\vskip 0.3cm
By defining
\begin{equation}
h(\alpha)=\frac{\alpha^2\Gamma(\alpha+3/2)}{(\alpha+1)\Gamma(\alpha+2)},
\label{z9}
\end{equation}
and collecting in $V_2$ and $V_5$ unknown constants the result, for 
the gluonic Pomeron contribution to the amplitude, is
\begin{equation}
A_g=\frac{-h(\alpha)}{\sin(\pi\alpha/2)}\left(\frac{-is}{s_0} \right 
)^{\alpha} [(\alpha-1)V_2(t)+(\alpha+1)V_5(t)]^2.
\label{z10}
\end{equation}
Equation (\ref{z10}) can be easily generalized to the case $a\neq b$. It 
is sufficient to substitute the term, within squared brackets in 
(\ref{z10}), with the product: $ [(\alpha-1) V_2^{(a)}(t)+(\alpha+1) 
V_5^{(a)}(t)]\cdot [(\alpha-1) V_2^{(b)}(t)+(\alpha+1) V_5^{(b)}(t)]$.
\vskip 0.3cm
In addition to well established properties of the amplitude, like Regge 
behaviour, a new feature appears. If a linear trajectory is adopted for 
the Pomeron, a term vanishing with $t$ is present  in (\ref{z10}) since
$V_2(t)$, as shown below, can be made regular at $t=0$. This 
term can have an important r\^ole in the calculation of the forward 
slope and in the dip region. At any rate, it 
shows that the $t$-dependence can be more involved than commonly 
believed. Corrections to eq.~(\ref{z10}) will give rise to other 
contributions that can be summarized as follows. 
\begin{itemize}
\item Non-leading contributions will appear in the 
amplitude because of the neglected terms in eq.~(\ref{z3}).
\item Since the integrals over $k_-$ and $k_+'$ comprise a region where 
these variables are not large, and hence do not correspond to the 
exchange of a Regge pole, the contribution of this region must be 
subtracted from eq.~(\ref{z10}). The most important term will appear from
the integration of the coefficient of $V_5(t)$ in eq.~(\ref{z6}), since 
it is multiplied by $\omega^2$. The contribution to the amplitude will 
be proportional to $s V_5^2(t)$.
\item More important will be the contribution due to the presence of 
quarks in the Pomeron. The coupling of the Pomeron to a quark in the 
hadron has the form~\cite{OT,GJ}
\begin{displaymath}
C_{\alpha}^+C_{\beta}^+\frac{\partial}{\partial P_{\beta}}\frac{\partial 
}{\partial Q_{\alpha}}(P\cdot Q)^{\alpha}
\end{displaymath}
where
\begin{displaymath}
C^+_{\alpha}(\frac{1}{2}\,\frac{1}{2}\,J)=(f_1 Q_{\alpha}+f_2\gamma_{\alpha})
\end{displaymath}
and an analogous expression for $C^+_{\beta}$. In the eikonal 
approximation, $Q_{\alpha}$ and $\gamma_{\alpha}$ give the same 
contribution and, for the amplitude, one obtains
\begin{equation}
A_q=\frac{\alpha (\alpha+1)\Gamma(\alpha+3/2)}{\sin(\pi\alpha/2) \Gamma
(\alpha+1)}\left( \frac{-is}{4}\right)^{\alpha}V_q^2(t).
\label{z11}
\end{equation}
with a $t$-dependence different from the one found in eq.~(\ref{z10}). 
\item Non-leading trajectories will give the same contribution, with the 
appropriate change in the trajectory function $\alpha(t)$, as in eq. 
(\ref{z11}).
\end{itemize} 

\section{A dipole Pomeron model}

The amplitude in eq.~(\ref{z10}) is an asymptotic estimate and, 
in order to preserve unitarity, it is preferable to keep the Pomeron intercept 
at one and adopt a dipole Pomeron model~\cite{LLJ} to account for rising 
cross sections. It is well possible to consider instead a supercritical 
Pomeron with an intercept slightly higher than one~\cite{BDL} but, as 
will be clear later, 
 the  formalism of Section {\bf 2} is particularly suited for a model where the 
Pomeron is a double pole in the J-plane. Eq.~(\ref{z10}) 
distinguishes clearly the functional dependence on $t$ and on $\alpha(t)$. 
Since the procedure to obtain a dipole Pomeron amounts to derive 
eq.~(\ref{z10}) with respect to $\alpha(t)$, the answer is 
unique~\footnote{A possible dependence of $g_1$ and $g_2$ on 
$\alpha$, as would happen for example in a dual model, does not change 
the final conclusions.}.
\vskip 0.3cm
Only the gluonic component of the dipole Pomeron will be considered in 
the following and, setting
\begin{equation}
W\equiv W(\alpha,t)=(\alpha-1)V_2(t)+(\alpha+1)V_5(t),
\label{z12}
\end{equation}
from eq.~(\ref{z10}) the imaginary and real part of the amplitude assume 
the form:
\begin{equation}
{\cal I}m A_g^{(d)}= \left(\frac{s}{s_0} \right)^{\alpha} 
W\left(h\left[\ln\left(\frac{s}{s_0}\right)W+2\frac{dW}{d\alpha}\right]+
\frac{dh}{d\alpha}W\right),
\label{z13}
\end{equation}
and
\begin{displaymath}
{\cal R}e A_g^{(d)}= -\left(\frac{s}{s_0} \right)^{\alpha} 
W\left(hW\left[\ln\left(\frac{s}{s_0}\right)\cot\left(\frac{\pi\alpha}{2}\right)
-\frac{\pi}{2\sin^2(\pi\alpha/2)} \right]\right. +
\end{displaymath}
\begin{equation}
+\left. \cot\left(\frac{\pi\alpha}{2}\right)\left[\frac{dh}{d\alpha}W+
2h\frac{dW}{d\alpha}\right]\right)
\label{z14}
\end{equation}
The trajectory can be chosen in the form $\alpha(t)=1+\alpha' t$, with
the conventional value for $\alpha'$, $\alpha'=0.25\;GeV^{-2}$. As a 
consequence of the derivation, with respect to $\alpha(t)$, 
factorization of residues is lost in eqs.~(\ref{z13}) and (\ref{z14}). 
The factorization breaking term is proportional to $d\ln W/d\alpha$.
\vskip 0.3cm
The total cross section
\begin{displaymath}  
\sigma_T=\frac{Im A^{(d)}(s,0)}{s},
\end{displaymath}
the differential cross section
\begin{displaymath}
\frac{d\sigma}{dt}=\frac{1}{16\pi s^2}\left|A^{(d)}(s,t) \right|^2,
\end{displaymath}
the forward slope and the ratio between the real and imaginary part of 
$A^{(d)}(s,t)$ can be derived from the total amplitude $A^{(d)}$ that will be
the sum of three terms
\begin{displaymath}
A^{(d)}=A_g^{(d)}+A_q^{(d)}+A_{n.l.},
\end{displaymath}
where $A_{n.l.}$ represents the contribution of non-leading 
trajectories, as simple poles, and $A_q^{(d)}$ is the derivative of 
eq.~(\ref{z11}) with respect to the Pomeron trajectory.
\vskip 0.3cm
Before trying to obtain more concrete informations from this model, it 
is important to spend a word on the integrals implied in the 
correspondences (\ref{z4}, \ref{z5}). At large $|t|$, perturbative QCD 
is expected to give the correct answer and the integration over the 
gluon propagator, that in this regime has the form 
$D(\vec{k}_{\perp}^2)\simeq 1/\vec{k}_{\perp}^2$, does not give rise to 
problematic results. However, for small $|t|$, divergencies will arise in 
the derivatives of $V_2(t)$ and $V_5(t)$; for example, in the Born 
approximation, two-gluon exchange results in a diverging slope at 
$t=0$~\cite{LR}.
First and second derivatives of the amplitude are related to the 
slope and curvature parameters, both have been measured experimentally 
in $p-\bar{p}$ scattering~\cite{E710}. The correct, but unknown, 
behaviour of the gluon propagator in the infrared must restore the 
physical properties of the scattering amplitude.
\vskip 0.3cm
It is possible to avoid the need of a precise representation for the non 
perturbative gluon propagators since  a superconvergence relation
exists for its discontinuity in Landau gauge~\cite{KN,RO}. 
Let $\sigma(k^2)$ be the discontinuity along the positive 
real $k^2$-axis of $D(k^2)$, the structure function of the transverse 
gauge propagator, $\pi\sigma(k^2)=Im D(k^2+i0)$. Then, if $N_f <10$ where 
$N_f$ is the number of flavours, the superconvergence relation 
\begin{displaymath}
\int_0^{\infty}\,d\lambda^2\,\sigma(\lambda^2)=0
\end{displaymath}
follows from the dispersion relation
\begin{displaymath}
D(k^2)=-\int_0^{\infty}\,d\lambda^2\,\frac{\sigma(\lambda^2)}{k^2-\lambda^2+
i\epsilon}.
\end{displaymath}
For large $|t|$, $\sigma(\lambda^2)\sim \delta(\lambda^2)$ reproduces the usual 
Feynman rule for the gauge propagator.  
Hence it is possible to add, to every troublesome integrals in eqs.~(\ref{z4}, 
\ref{z5}) a term of the form
\begin{equation}
R(t)\int\int\,\sigma(x)\sigma(y)\,dx\,dy=0
\label{z15}
\end{equation}
and regularize the divergence at $t=0$. This procedure does not introduce free parameters 
since the value of all integrals is fixed in the perturbative region 
and, according to~\cite{LR}, confinement effects become sensible only at 
quite small $k^2$, for example $|\vec{k}_{\perp}|\sim 2m_{\pi}$. In this 
narrow $t$ interval, linear, or quadratic, extrapolations should be 
possible with suitable matching conditions, at least from a 
phenomenological point of view~\cite{DL}.
It is assumed, in the following that this regularization takes into account 
all nonpertubative effects.

\section{Comparison with the standard approach}

Consider now proton-antiproton elastic scattering as 
a typical process that, at very high energy, can be described with the 
help of eqs.~(\ref{z12}), (\ref{z13}) and (\ref{z14}). The limitation to 
high energies is due to the neglect in the following of non leading 
terms:  mesonic trajectories and corrections to the Pomeron 
contribution.
\vskip 0.3cm
With the conventional Pomeron trajectory, $\alpha_P=1+\alpha'\;t$ with 
$\alpha'=0.25\;GeV^{-2}$, eq.~(\ref{z12}) gives
\begin{equation}
W=\frac{1}{4}\,t\,(V_2+V_5)+2 V_5,
\label{z16}
\end{equation}
while
\begin{equation}
\frac{dW}{d\alpha}=V_2+V_5,
\label{z17}
\end{equation}
where $V_2$ and $V_5$ are unknown functions of $t$.
Since 
\begin{displaymath}
h(\alpha)|_{t=0}=\frac{3\sqrt{\pi}}{16},\;\;\;\;\;\frac{d\ln 
h}{d\alpha}\left|_{t=0}=\frac{8}{3}-2\ln 2 \right.,
\end{displaymath}
the total cross section can be written as
\begin{equation} 
\sigma_T=\frac{3\sqrt{\pi}}{4s_0} V_5^2(0) \left[ \ln\left( 
\frac{s}{s_0} \right) +\frac{V_2(0)}{V_5(0)}+\frac{11}{3}-2\ln 2 
\right].
\label{z18}
\end{equation}
The differential cross section (in $mb^2$) has the form
\begin{displaymath}
\frac{d\sigma}{dt}=\frac{1}{16\pi 
s^2}\left(\frac{s}{s_0}\right)^{2\alpha} \frac{h^2 
W^4}{\sin^2(\pi\alpha/2)}\times
\end{displaymath}
\begin{equation}
\times\left(\left[\ln\left(\frac{s}{s_0}\right)+\frac{d\ln(hW^2)}{d\alpha}- 
\frac{\pi}{2}\cot\left(\frac{\pi\alpha}{2}\right) 
\right]^2+\frac{\pi^2}{4} \right)
\label{z19}
\end{equation}
and, from eq.~(9), it is easy to obtain
\begin{displaymath}
\frac{d\ln h}{d\alpha}=\frac{2}{\alpha (\alpha +1)}-\psi(1+\alpha)+ 
\psi(\alpha+3/2),
\end{displaymath}
where $\psi(z)$ is the logarithmic derivative of the gamma function. 

\begin{figure}[!hb]
\centering
\includegraphics[width=0.9\textwidth]{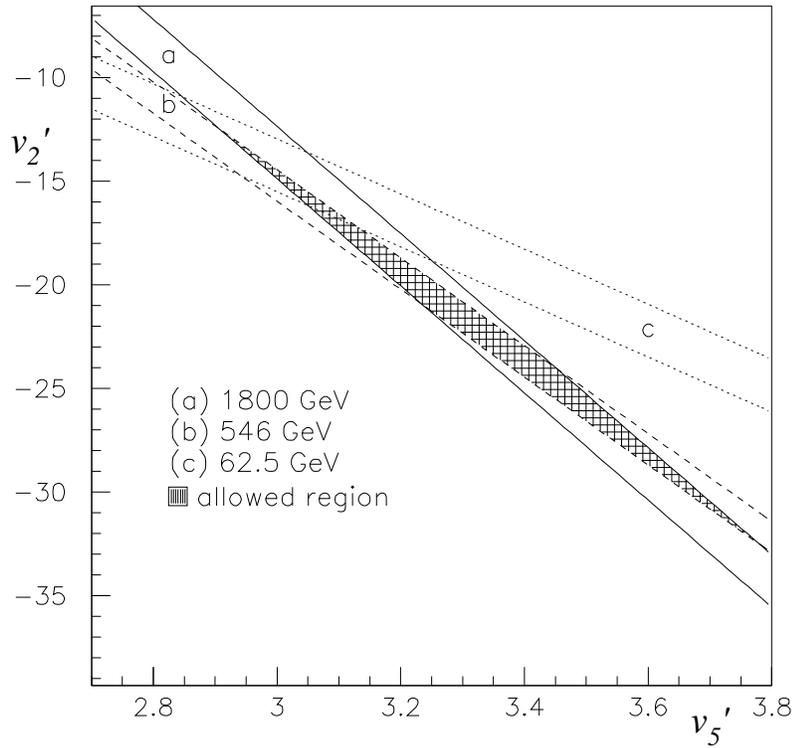}
\caption[]{\small Allowed region in the $(v_5{}',v_2{}')$ plane. Continuous 
lines are determined from the slope, with errors, at $\sqrt{s}=1800\;GeV$ 
, dashed lines refer to $\sqrt{s}=546\;GeV$ and dotted lines to 
$\sqrt{s}=62.5\;GeV$. Data are from~\cite{CDF,ISR2}.}
\label{fig4}
\end{figure}

The slope in the forward direction, in this approach, is a cumbersome 
expression that presents, however, many interesting features. As will be 
shown later, there is in fact the possibility to overcome discrepancies, present 
in the standard approach, appearing when the experimental slope is 
fitted at different energies. In order to have a compact form for this 
observable, it is convenient to introduce the following  notation:
\begin{displaymath}
f(s)=\ln\left( \frac{s}{s_0} \right)+\frac{8}{3}-2\ln 2,
\end{displaymath}
and 
$V_5{}'(0)/V_5(0)=v_5{}',\;V_2(0)/V_5(0)=v_2,\;V_2{}'(0)/V_5(0)=v_2{}'$. 
With these definitions, the forward slope is
\begin{displaymath}
b\equiv\left. 
\frac{d}{dt}\left(\ln\frac{d\sigma}{dt}\right)\right|_{t=0}= 
2\alpha'f(s)+\frac{v_2+1}{2}+4 v_5{}'+
\end{displaymath}
\begin{equation}
+2\left[v_2{}'-v_5{}'v_2-\frac{1}{8}(v_2+1)^2+\frac{\alpha'}{6} 
\left(\frac{7\pi^2}{2}-\frac{89}{3}\right)\right]\frac{f(s)+v_2+1}{(f
(s)+v_2+1)^2+\pi^2/4}.
\label{z20}
\end{equation}
\vskip 0.3cm
In order to see how problems in fitting experimental data can be 
removed in a specific example, let us consider $p-\bar{p}$ elastic 
scattering at the Tevatron~\cite{CDF}. Measurements of the elastic 
slope, near $t=0$, give $b=15.35\pm 0.18\;GeV^{-2}$, at 
$\sqrt{s_1}=546\;GeV$, and $b=16.98\pm 0.24\;GeV^{-2}$ at $\sqrt{s_2} = 
1800\;GeV$. Assuming an $s$-dependence of the slope $b=b_0+2\alpha' 
\ln(s/s_0)$, the data at these energies, where only the Pomeron 
probably contributes, give $\alpha'=0.34\pm 0.07\;GeV^{-2}$ to be 
compared with a lower value when the data at the 
CERN-ISR~\cite{ISR1,ISR2} are included in the fit. From $546\;GeV$ to 
$1800\;GeV$, the total cross section increases by $18.8\pm 2.5\;mb$, 
according to~\cite{CDF}. Then eq.~(\ref{z18}) implies that 
$V_5^2(0)/s_0=5.9\pm 0.8\;mb$ while the measured values of the total cross 
section at these energies give
\begin{displaymath}
v_2-\ln s_0=-7.0\pm 1.6.
\end{displaymath}
By keeping the central value for $(v_2-\ln s_0)$ and fixing the value of 
$s_0$, for example~\footnote{Remember that, in this model, $s_0> 
4\;GeV^2$ (see Section {\bf 2}).}  $s_0=9\;GeV^2$, it is possible to find 
the allowed region, in the plane $(v_5{}',v_2{}')$, determined from the
experimental slopes, and their errors, at the energies $\sqrt{s_1}$ and 
$\sqrt{s_2}$. In this calculation, $\alpha'$ is always $0.25\;GeV^{-2}$.
Figure 4 shows the region where $b$, in eq.~(\ref{z20}), 
satisfies the experimental bounds given by~\cite{CDF}. If $s_0$ is 
increased, above $9\;GeV^2$, the parallelogram representing the allowed 
region moves down and to the right, in the $(v_5{}',\,v_2{}')$ plane, 
preserving its form. In this figure, 
also the boundary determined by the ISR data~\cite{ISR2} at $\sqrt{s}= 
62.5\;GeV$ is shown. This can be considered as an extreme example, since
I am not aware of other parametrizations, within the Regge framework,
that succeed in reproducing the total cross sections, measured 
by~\cite{CDF}, at both energies. Usually the cross section at $1800\;GeV$ 
is underestimated.
\vskip 0.3cm
The term proportional to $t$, in the vertex, has now an important r\^ole 
and makes the determination of the slope somewhat independent on the 
actual value of the total or the differential cross section in the 
forward direction. It is plausible that, for $t$ different from zero, this
term could help in explaining non trivial 
properties of the forward cone. In order to substantiate this belief, it is 
interesting to consider the real and imaginary part of the amplitude, 
eqs.~(\ref{z13}) and~(\ref{z14}).
There are both theoretical~\cite{AM} and phenomenological~\cite{PF,DGM} 
reasons for the presence of a zero in the real part of the even signature 
amplitude near $t=0$. Looking at eq.~(\ref{z14}), this requirement
can be written in the form
\begin{equation}
\ln\left(\frac{s}{s_0}\right) +\frac{d\ln h}{d\alpha}-\frac{\pi}{ 
\sin(\pi\alpha)}=-\frac{d\ln W}{d\alpha},
\label{z21}
\end{equation}
since $W$ cannot vanish. The l.h.s. in eq.~(\ref{z21}) has a zero in $t$ 
that depends on $s$: it vanishes for $t=-0.34\;GeV^2$ at $546\;GeV$ and 
for $t=-0.28\;GeV^2$ at $1800\;GeV$, $(s_0=9\;GeV^2)$. These values are 
near to the finding of~\cite{DGM} and corrections, due to the presence 
of the r.h.s. in eq.~(\ref{z21}), are small if the ratio $V_2(t)/V_5(t)$ 
does not change too much from its value at $t=0$, $v_2$. An increase of 
this ratio with $|t|$ shifts the position of the zeros towards smaller 
values of $|t|$. Predictions for the region of larger values of $|t|$ 
are speculative since other contributions, for example the odderon, will 
become important. Notice however that, if the ratio $V_2(t)/V_5(t)$ 
continues to increase with $|t|$, also the imaginary part of the 
amplitude can vanish. According to~\cite{PF}, the zeros of the imaginary 
part of the amplitude, together with the above results for the real 
part, are important in the description of the dip observed in the 
differential cross section.
\vskip 0.3cm
It is well possible the an analogous mechanism could be relevant
in the case of the electroproduction of heavy vector mesons at 
HERA. For example, the flattening of the slope for the production of
$J/\psi$ can be explained in this approach without requiring a drastic
change in the slope of the Pomeron trajectory. 
According to the generalization proposed after eq.~(\ref{z10}), the 
differential cross section for the photoproduction of the $J/\psi$ 
is obtained from eq.~(\ref{z19}) with the substitution 
\begin{displaymath}
W \rightarrow \sqrt{UW}
\end{displaymath}
where the new vertex $U(t)$ has the form given in eq.~(\ref{z16}), but 
refers to the vertex $\gamma$-Pomeron-$J/\psi$. An attempt along this 
line has been considered in ref.~\cite{FJP}. In the case of 
electroproduction, this vertex will be a function of $t$ and $Q^2$, 
where $q^2=-Q^2$ is the square of the fourmomentum of the off-shell 
photon. Data for the diffractive production of vector mesons have been 
published by H1~\cite{H1} and ZEUS~\cite{ZEUS} Collaborations. The 
scarcity of experimental data and their large errors, for the cases
of interest here, where only the Pomeron is exchanged, does not allow
an analysis similar to the one performed for the $p-\bar{p}$ case.
It is plausible, however, that the greater flexibility reached in this
model will help in accounting for the variation of the slope with 
energy.

\section{Concluding remarks}

The proposed approach to the hadronic scattering at asymptotic energies 
regards the Pomeron propagator as a Regge dipole, while its coupling to 
the quarks in the hadrons reflects the Pomeron structure in terms of 
gluons and quarks. Covariant reggeization determines the general form of 
the scattering amplitude for the interaction of two off-shell gluons 
when the Pomeron is exchanged. Since the internal colour structure of 
the hadrons has not been specified, only general properties of the 
amplitude for the hadronic process, imposed by the method of 
reggeization, can be derived.
\vskip 0.3cm
From the operative point of view, the main difference with the standard 
approach consists in the appearance of a new term in the amplitude, 
that, if a linear trajectory is adopted for the Pomeron, vanishes 
linearly in the forward direction. The importance of this term in the 
description of experimental data, especially when the slope in the 
forward direction is considered, is shown in the specific example of 
$\bar{p}-p$ scattering.  
The Pomeron trajectory can be fixed, once for all, in the 
description of the forward slopes at different energies. In the dipole 
Pomeron model~\cite{LLJ}, adopted in this paper, the increase of 
the total cross section can be obtained with a unity intercept for the 
Pomeron trajectory, while its slope coincides with the conventional 
value of $0.25\;GeV^{-2}$. Corrections due to the presence of quarks in the Pomeron and 
non leading contributions have been explicitly evaluated in Section {\bf 
2}.
The case of photon induced processes is also briefly discussed.
\vskip 0.3cm
It would be tempting to parametrize the functions appearing in the 
vertex (see eq.~(\ref{z16})) in terms of exponentials, and/or  
ratios of polynomials, and obtain an amplitude where the number of free 
parameters is comparable to other parametrizations. From a 
phenomenological point of view, this approach could be adequate.
However, in my opinion, the correct way to increase 
the predictability of the model must start from the explicit calculation 
of the vertex with different forms of the hadron wave functions, 
available in the literature. The comparison of the result with the 
experimental data requires, in addition, the knowledge of the Pomeron 
couplings to off-shell gluons. A hint for the latter couplings could 
come from the (skewed) parton distribution functions in the Pomeron.
\vskip 0.3cm
The limits of the model are set from the chosen framework: 
Regge exchange and momentum space technique are the main ingredients of 
this calculation. Both are supposed to describe correctly the scattering 
amplitude only in the small $|t|$ region. 
The choice of a different form for the Regge trajectories~\cite{BCJ} 
could provide a smooth interpolation between soft and hard behaviour of 
the scattering amplitude, from an exponential decrease to a power law in 
in $t$~\cite{BCJ}. It would be quite interesting to study, within the present 
model, the effect of a Pomeron trajectory, with a two-pion square root 
threshold, on the differential cross section~\cite{JSS}. This calculation, 
that could also explain the presence of kinks in the differential cross section well 
before the dip, will be considered elsewhere.

\vspace{0.5cm}
{\bf \large Acknowledgments}

I thank Laszlo Jenkovszky for numerous and stimulating discussions on 
the Pomeron.

\end{document}